\documentclass[runningheads]{llncs}
\usepackage[T1]{fontenc}
\usepackage{graphicx,float,verbatim,hyperref}
\begin{document}
\title{The Self-Aware Body: A User-Centered Framework for Designing Therapeutic Sonic Interactions}
\titlerunning{The Self-Aware Body: Therapeutic Sonic Interaction Framework}
%
\author{Prithvi Ravi Kantan\inst{1}\orcidID{0000-0001-5618-4265} \and
Sofia Dahl\inst{1}\orcidID{0000-0002-0687-7810} \and
Erika G. Spaich\inst{2}\orcidID{0000-0003-1275-4064}}
\authorrunning{P.R. Kantan et al.}
\institute{Department of Architecture, Design and Media Technology, Aalborg University, Copenhagen, Denmark\\
\email{[prka,sof]@create.aau.dk} \and
Department of Health Science and Technology, Aalborg University, Aalborg, Denmark\\
\email{espaich@hst.aau.dk}}
\maketitle              
\begin{abstract}
This chapter presents a framework for designing therapeutic sonic interaction technologies, with a focus on movement sonification: the real-time conversion of bodily motion into sound that serves as feedback during motor rehabilitation. Despite growing evidence for their effectiveness, technologies implementing movement sonification are yet to be systematically adopted as part of clinical practice, potentially due to a lack of standardized development methodologies as well as inadequate integration of clinical stakeholder perspectives into interaction design. The framework addresses these barriers through three interconnected contributions. The first is a conceptual reframing of the design task as the calibration of sonic variability to the perceptual affordances of the listener and the demands of the clinical context. The second is a practical design platform inspired by professional audio mixing workflows, which imposes a structured and learnable signal-flow architecture on the interaction design process and enables rapid iterative exploration. The third is a user-centered development methodology adapted from healthcare intervention science, which grounds design decisions in engagement with the clinicians and patients who will use the resulting systems. The HearWalk biofeedback system for hemiparetic gait rehabilitation illustrates the framework, and the chapter concludes by examining where large language models and AI tools can meaningfully assist each stage of this design process, as well as where human clinical and perceptual expertise remains irreplaceable.
\end{abstract}
\keywords{movement sonification \and auditory biofeedback \and motor rehabilitation \and sonic interaction design \and user-centered design}

\section{Introduction to Movement Sonification as Auditory Biofeedback}

Sound pervades human experience in ways that are easy to overlook: it orients us in space, coordinates our movement with others, marks the rhythms of shared activity, and continuously colors the texture of everyday life with information about the world around us. Yet its influence is not merely contextual or informational. Across human cultures and throughout recorded history, organized sound---music---has been recognized as capable of altering mood, sustaining attention, modulating pain, and eliciting movement in ways that have proven consistent enough to ground formalized therapeutic practice \cite{zaatar2024transformative,rebecchini2021music,maes2016_3MO}. Music can modulate activity in brain structures governing emotion, reward, and autonomic arousal \cite{koelsch2014brain}, and decades of clinical research have demonstrated that deliberate use of sound and music can produce meaningful therapeutic effects \cite{demetriou2026scoping,schaffert2019review,bradt2021cochrane}. Music therapy, or the clinical use of music-based interventions delivered by trained therapists, has accumulated robust evidence for its effects on psychological wellbeing, pain perception, and quality of life across a wide range of conditions including depression, dementia, cancer, and chronic pain \cite{thaut2014nmtBook,bradt2021cochrane}. Neurologic music therapy has formalized this potential into structured, evidence-based techniques that directly harness the brain's auditory-motor coupling, i.e., the extensive neural connectivity between auditory and motor systems, to support gait training, speech rehabilitation, and upper limb recovery following neurological injury \cite{thaut2014nmtBook}. What unites these applications, beneath their considerable surface differences, is the core notion that sound can function as an active therapeutic medium, capable of influencing how people sense, regulate, and modify their own bodies.

The emergence of digital technology has opened a new chapter in this story. Real-time computational systems can now generate and adapt sound in response to a person's body or behavior, enabling forms of sonic interaction that go beyond the affordances of the physical domain in a conventional sense \cite{effenberg2005movement}. Generative music systems, immersive sonic environments, and sensor-driven audio tools have been explored as instruments for anxiety reduction, attention regulation, and embodied self-awareness \cite{demetriou2026scoping}. In these applications, the real-time responsiveness of the sonic medium allows the creation of experiences that adapt dynamically to the individual. This positions interactive digital sonic systems as a promising frontier in health and rehabilitation: one in which the therapeutic affordances of sound are extended and personalized through real-time interactivity \cite{sigrist2013augmented,guerra2020soniPhysio}. As such, this chapter frames a \emph{self-aware body} as one that can monitor and modify its own movement through the auditory channel and provides a design framework targeting clinical rehabilitation contexts. The design challenge, as the following sections develop, is to engineer sonic responses such that patients can use the auditory channel to perceive and regulate their own movement without the sound overwhelming the attentional resources needed for the motor task itself.

Within this landscape, one application has attracted particular attention for its potential in physical rehabilitation: \emph{movement sonification}, or the real-time translation of bodily motion into sound \cite{effenberg2005movement,schaffert2019review}. In a movement sonification system, sensors capture kinematic data such as measurements of joint angles, velocities, and accelerations during movement, which an algorithm converts in real time into a continuously or phasically evolving sound signal \cite{effenberg2005movement}. This can be termed \emph{auditory biofeedback} (ABF), i.e., information about how one is moving, delivered through the auditory channel in real time \cite{kos2018bmbf}. Where conventional rehabilitation typically relies on visual displays, haptic feedback actuators, or the verbal cues of a therapist, ABF creates an additional perceptual channel with several distinctive advantages: it leaves the patient's visual attention free for the movement task itself \cite{sigrist2013augmented}; it offers fine temporal resolution for communicating subtle kinematic details \cite{eddins1995temporal}; and evidence from comparative studies suggests that ABF may promote better long-term motor retention than its visual counterpart, potentially because the feedback strengthens the coupling between auditory and proprioceptive systems that underpins automatized movement \cite{sigrist2013augmented,ma2016balance}. The underlying neuroscience further motivates this approach, as extensive connectivity between auditory and motor regions of the brain means that sound has direct pathways into the circuits governing movement \cite{lahav2007action}. This makes the auditory channel a physiologically grounded and well-motivated route for delivering movement information in rehabilitation.

The standard technical vehicle for implementing ABF is \emph{parameter mapping sonification}: movement variables derived from sensor data are linked in real time to audio parameters such as pitch, amplitude, timbre, and spectral content through a mapping function, so that the sound evolves continuously as movement unfolds \cite{grond2011parameter,hermann2008taxonomy}. Defining this mapping requires more than specifying what kinematic information is encoded. The resulting sound must remain semantically interpretable throughout a clinical session without consuming the attentional resources the patient needs for the movement task, and aesthetically sustainable across weeks of repetitive training. Designs that prioritize signal clarity above all else, such as mapping a kinematic variable directly to a sine wave frequency, for instance, consistently prove perceptually sterile and motivationally inert when tested with actual users \cite{dyer2016sonification,kramer2010sonification}. The field has responded by grounding design in embodied cognition and the notion of \emph{conceptual metaphors}: sounds that engage everyday sensorimotor listening schemas allow patients to interpret feedback through familiar perceptual experience rather than explicit decoding, reducing cognitive overhead and sustaining engagement over time \cite{leman2018embodied,roddy2020mapping}. Mapping the angular velocities of the legs to the amplitude and spectral content of water wading sounds, for instance, produces feedback that stroke patients interpret intuitively and without instruction, by analogy with the heard experience of walking through water \cite{kantan2023ecological,kantan2024wading}. Sound design and motor learning have largely been pursued by separate research communities, and how feedback sounds actually communicate a kinematic variable to a listener who is simultaneously managing a cognitively demanding movement task has received far less scrutiny than the motor outcomes themselves \cite{bevilacqua2016sensorimotor}.

The promise of movement sonification is supported by a growing body of experimental work. Reviews of the field report improved motor learning, increased motivation, and meaningful sensorimotor effects across diverse applications including stroke rehabilitation, gait training or balance recovery in neurological/orthopedic patient populations, and sports coaching \cite{schaffert2019review,sigrist2013augmented}. Yet, despite these encouraging findings, movement sonification has not achieved meaningful clinical adoption \cite{guerra2020soniPhysio,nown2022mapping}. The reasons for this reflect three interconnected challenges. The first is conceptual: the field lacks a principled framework for understanding what makes sonic feedback perceptually meaningful rather than merely informationally correct \cite{dyer2015movementSoni}. Effective auditory feedback must not only encode kinematic variables precisely but also remain semantically interpretable and aesthetically sustainable across the repeated sessions that rehabilitation demands, which represents a tension between signal fidelity and perceptual richness. The second is methodological: the clinicians and patients who must ultimately use these systems have rarely been meaningfully involved in their development, resulting in designs that may be technically functional but clinically unworkable in practice \cite{guerra2020soniPhysio,wang2020faster}. Meaningful stakeholder engagement requires a development process built on rapid iterative prototyping. This in turn exposes the third challenge: sonification design has typically been carried out in general-purpose audio programming environments or through custom scripts in tools built for audio prototyping or signal processing in isolation, not for the iterative process that clinical development requires. In practice, a sonification design is realized in the form of one or more \emph{mappings}, which are essentially causal links between movement-related data variables and audio synthesis/processing parameters that make the experience of movement-sound interaction possible. Modifying a mapping means editing code or rewiring a `patch' in a visual programming environment (e.g. Max/MSP); evaluating the result of the modification means re-running the system rather than adjusting a parameter and listening immediately. Given the inherently iterative nature of this process, it is crucial that sonification design platforms are conducive to rapid real-time modification of mappings with minimal temporal delay between mapping modifications and their perceived impact on the sonic interaction itself.

This chapter presents a framework for therapeutic movement sonification built around three contributions. The first reframes the conceptual problem: rather than asking how movement data should be faithfully encoded into sound, it reconceives the design task as the calibration of sonic variability to the perceptual affordances of the target listener. This creates an explicit distinction between the informational content of the feedback sound signal and the information that a listener in a clinical context can actually use, and develops the vocabulary needed to navigate it. The second adapts a framework from healthcare intervention science into a structured process for engaging physiotherapists and patients iteratively throughout development, grounding design decisions in the operational realities of clinical practice rather than assumed requirements. The third introduces a design platform modelled on the architecture of the audio mixing console; in place of code-level patch editing and offline testing cycles, it provides a structured, real-time parameter control environment that operates with live sensor data and produces a transparent, inspectable design artifact at every stage of development. The HearWalk system, a clinically evaluated ABF tool for hemiparetic gait rehabilitation \cite{kantan2024water}, illustrates the three contributions applied in full. The chapter also examines where large language models can meaningfully assist this design process, and where the irreducibility of human clinical and perceptual expertise sets limits on their contribution. 


\section{The Signal-Noise Paradox: Rethinking Sonic Meaning}

Even with an appropriate sonic metaphor, realizing a usable mapping requires a pipeline of processing operations between the raw sensor output and the audio parameter: normalization to a standard range, smoothing to suppress sensor noise, nonlinear transformation to bring relevant movement variations into a perceptually sensitive region of the parameter's range, and scaling to the target synthesis parameter \cite{kantan2024wading,grond2011parameter}. The configuration of this pipeline interacts with both the target movement and the sound model chosen as the feedback medium, and this last dependency is less obvious but consequential. All sounds except static synthesized waveforms exhibit varying degrees of short-term spectrotemporal variation: natural sounds like wind and water exhibit stochastic fluctuations in amplitude and frequency content, and most music is characterized by rhythmic and harmonic variation in time. The same applies to the sound models used in movement sonification, wherein every model introduces variation of its own, independent of the user's movement. The audibility of movement-sound correspondence therefore depends on how movement-induced changes in the audio signal relate to this intrinsic variation. Consider a sonification based on generative multitrack music whose rhythm and harmony spontaneously evolve according to its internal algorithm, and where movement during balance training controls one specific aspect of the music, such as its performance dynamics. Such a sonification may communicate movement information clearly during vigorous, large-excursion movement, but become perceptually opaque when applied to the subtler bodily excursions characteristic of static balance training. This is not because the mapping is incorrect, but because movement-driven sound changes are too subtle to distinguish from the sound model's own natural evolution. Calibrating the relationship between the movement information to be conveyed and the sound model's autonomous sonic variation is the central conceptual challenge this section addresses, using the concept of the \emph{noise profile} to give the problem a workable vocabulary.

A basic insight guiding the framework concerns the relationship between informational `purity' and perceptual meaning in movement sonification. Consider a simple scenario: a single numerical variable such as the angular velocity of a patient's thigh during walking is to be sonified continuously in real time. The most `noise-free' implementation maps this variable directly to a sine wave frequency: as the leg moves faster, the pitch rises; as it slows, the pitch falls. Nothing else about the sound changes, not its loudness, not its timbre, not its spectral texture. The movement variable occupies exactly one acoustic dimension, with all others held constant. This makes the sine wave a maximally pure encoding, as the sound carries precisely one piece of information and introduces no variation of its own. From an information-theoretic perspective, this is optimal. Yet such designs consistently prove perceptually sterile, cognitively fatiguing, and semantically impoverished when tested with actual users \cite{dyer2016sonification,kramer2010sonification}. In this case, the \emph{most faithful} representation of the underlying information is also the \emph{least perceptually useful}. The path toward more meaningful feedback requires moving to richer sound models such as real-life sounds, rhythmic textures, resonant physical objects that engage existing sensorimotor associations and make feedback interpretable without deliberate decoding. But this gain in meaning is inseparable from a loss in the informational purity of the output signal. A richer sound model has its own internal dynamics: timbral evolution and/or amplitude fluctuation that occurs regardless of what the user's body is doing. From a signal-theoretic standpoint, these internal sound dynamics constitute `noise': variation in the output that cannot be attributed to the movement variable being communicated. The critical point is that this is not a correctable design flaw but an unavoidable entailment of choosing an expressive sound. Injecting meaning and injecting noise are two sides of the same coin: any sound model rich enough to carry perceptual meaning is, by definition, one with autonomous variation, and that variation is simultaneously what makes the feedback engaging. Ensuring that this variation does not obscure the movement-driven signal is therefore the core design challenge, i.e., calibrating sound models so that their intrinsic dynamics enrich rather than overwhelm what the movement is communicating. 

Our perceptual study comparing movement-sound congruence across model types provides empirical grounding for this framing of the core design challenge \cite{kantan2026beyond}. The study contrasted a deterministic sonification model (the sine wave being the limiting case of a noise-minimal baseline) with generative music models that produce temporally evolving musical structures driven by movement parameters while maintaining some autonomous variation. The deterministic model produced consistently high congruence ratings regardless of movement amplitude and frequency, confirming its status as a noise-minimal baseline. The generative models showed strongly movement-dependent behavior: congruence ratings were substantially modulated by movement amplitude and frequency, with small or slow movements yielding low perceived correspondence with the sound because the movement-induced sound changes were perceptually indistinguishable from the model's own autonomous dynamics. In the context of movement sonification, this work proposed the term \emph{noise profile} to describe the characteristic pattern of variation a sound model introduces independently of user input, against which movement-driven changes must be perceptible to register as clear variations in the biofeedback stimulus. Simple deterministic models have low-amplitude, time-invariant noise profiles; generative music models have richer, temporally and spectrally complex ones. The practical implication is that sound model choice cannot be decoupled from the movements being sonified; a sound model calibrated for large-excursion bodily movements may not be suited to conveying smaller movements, because the movement-driven signal cannot be heard against the model's intrinsic variation \cite{kantan2026beyond}. Conversely, a deterministic sine wave faithfully encodes subtle kinematic variations but fails on perceptual and motivational grounds over extended training sessions. The design space is therefore not a simple axis from noise-free to noise-rich: it is a multidimensional landscape that the designer must shape based on the movement type, patient population, patient preferences, session duration, and the specific impairments being targeted.

This frames the design task as one of calibration, i.e., tuning the relationship between that variation and the movement-driven signal for the specific movements, patient population, and clinical context in question. A model too far toward the noise-minimal end will most likely fail on meaningfulness-related and motivational grounds; one whose noise profile overwhelms the movement-dependent signal will most likely fail on functional grounds. The viable design region lies between these poles, in that it is expressive enough to sustain engagement over repeated sessions and simultaneously transparent enough for movement-sound correspondence to remain perceptible across the full range of movement patterns the target population presents. Reaching that region cannot be achieved through theoretical derivation alone and requires iterative empirical testing with real movement data. The challenge is compounded in therapeutic contexts by the nature of the movements involved: for instance, impaired walking is lower in amplitude, more variable, and often asymmetric in ways that healthy-user approximations do not capture. Any sonification design must accommodate movement characteristics and their interpersonal variability in order to be clinically viable. As such, sonification tools and design methods must be tailored to facilitate the realization of such designs, and the subsequent sections aim to provide a concrete and practical framework concerning precisely this.

\section{A Versatile Technological Platform for Constrained Exploration}

\subsection{Therapeutic Sonification as a Complex Intervention}

When sonification is used to promote motor recovery in patients with movement impairments, it formally acquires the status of a medical intervention: a deliberate activity undertaken with the objective of improving health by restoring function lost through disease or injury \cite{ross2015types}. More specifically, it qualifies as a \emph{complex intervention}, defined as one having ``several interacting components that impact the length and complexity of the causal chain from intervention to outcome'' \cite{craig2008developing}. Post-stroke motor rehabilitation is an archetypal complex intervention \cite{wade2005describing}, and augmenting it with ABF technology adds further layers of complexity relating to hardware reliability, software usability, physiotherapist training requirements, and the need to individualize feedback parameters to each patient's movement impairments and therapeutic goals. Framing ABF-augmented rehabilitation as a complex intervention is not merely a matter of terminology; it directly affects how the development process should be structured, what constitutes adequate evidence, and what counts as a meaningful step toward clinical adoption.

Several frameworks have been proposed to provide guidance on developing, evaluating, and implementing complex medical interventions rigorously. The Medical Research Council (MRC) guidelines \cite{craig2008developing} and their expanded version by Bleijenberg and colleagues \cite{bleijenberg2018mrc} describe a development process that is rationale-focused, iterative, and rooted in existing evidence and theory before any empirical testing begins. Wang and colleagues \cite{wang2020faster} adapted these principles alongside modern user-centered design methods to produce the FASTER framework, which is specifically tailored to technology-based interventions in rehabilitation and disability. FASTER organizes the development and evidence generation process into: \emph{Development}, \emph{Progressive Usability and Feasibility Evaluation}, and \emph{Scaled Evaluation and Implementation}. The crucial feature is that the first two are highly iterative and involve continuous engagement with clinicians, patients, and other stakeholders through rapid prototyping, think-aloud testing, focus groups, and small-scale feasibility studies. The emphasis on rapid prototyping is particularly relevant to ABF. Because sonification designs are perceptually opaque until actually heard and interacted with, describing a proposed feedback design to clinicians and patients without a working prototype produces little actionable feedback. Working prototypes are hence the unit of communication, and the speed with which they can be produced and modified directly determines how efficiently the iterative development cycle can run. This is precisely where the absence of adequate design tooling bites hardest: a development framework that depends on rapid iteration but has no tooling support for rapid iteration is practically difficult to execute.

User-centered design (UCD) \cite{dabbs2009user}, a methodological cornerstone of the FASTER framework, is an iterative model in which designers focus on user needs rather than assumed requirements throughout development \cite{gould1985designing,or2022human}. In healthcare, this requirement is more demanding than in most domains: systems must fit into existing clinical workflows, accommodate wide variation in patient abilities, cognitive states, and needs and be operable by clinicians with no technical background and limited time for training \cite{chandran2020exploring}. Critically, UCD depends on tangible prototypes rather than verbal descriptions or paper specifications. It is only when clinicians and patients can hear, interact with, and respond to a working system that the feedback produced is grounded enough to drive meaningful design decisions \cite{dabbs2009user}. This is the point at which the tooling gap identified previously becomes most consequential. The speed and ease with which a prototype can be modified and re-evaluated determine whether clinician and patient feedback can be gathered and acted upon within the practical constraints of a development cycle. Rapid iteration is thus the core mechanism through which UCD actually functions. The framework presented in this chapter addresses this need directly, and the next two sections describe the conceptual and technical contributions 
(a design platform and a structured development process) that make true rapid iteration with clinicians and patients practically achievable.

\subsection{Mix-N-Map: From Mixing to Sonification Design}

Audio mixing is the process of combining and processing multiple audio tracks to produce a final mix \cite[Chapter 1]{izhaki2017mixing}. It has developed over several decades from an intuition-driven craft into a substantial technical discipline, supported by a mature body of tools, conceptual vocabulary, and transferable heuristics. What enabled this development was not primarily the sophistication of individual processors, but the emergence of the multitrack mixing console. By imposing a consistent structure on the workflow, the console created a shared frame within which engineers could develop expertise, communicate decisions to collaborators, and accumulate transferable skill over years of practice with real material \cite{izhaki2017mixing,kantan2024mixing}. Rather than confronting an engineer with unlimited possibilities for how to treat a sound, the console presents a finite set of meaningful decisions: how loud should this element be relative to the others; how bright or warm in tone; how much spatial depth and reverberation? Each has a dedicated control, consistent across every channel and every project, so the engineer never has to invent a workflow from scratch. The platform has already decomposed the problem into a fixed set of audible dimensions, directing attention toward the choices that matter and away from those that do not. It is this structured narrowing that makes skill development cumulative: the same decisions recur in recognizable form across projects, and a solution found in one context may be transferable to the next.

The structural logic of the mixing console transfers directly to the challenge of sonification design \cite{kantan2024mixing}. Both are fundamentally iterative search processes, where the designer applies a sequence of adjustments to a raw input signal and then evaluates the result by ear, refining the settings until the output meets the application's requirements. In both cases, the task is to translate a domain-specific goal (a musical vision or clinical criterion) into concrete parameter choices through a structured set of available operations. In the sonification context, these operations address questions such as: 
\begin{itemize}
    \item How sensitive should the sound be to a given movement, so that clinically relevant variations in gait are audible without sensor noise overwhelming them?
    \item Should rapid fluctuations be smoothed so that the feedback reflects the overall quality of a movement rather than reacting to each momentary measurement?
    \item How should the relationship between movement size and sound change be shaped to remain informative across the movement range of a patient?
    \item How quickly should the sound build and fade in response to a movement event?
\end{itemize}  
Each of the above aspects can be manipulated using a sequence of elementary signal processing operations such as filtering, gain control, and nonlinear or other time-variant processing, which directly correspond to the mixer's faders and knobs \cite[Chapter 14]{huber2013modern}. What transfers from the mixing console is not just these individual tools but the organizing architecture itself: a consistent signal-flow structure in which design decisions are visible, inspectable at a glance, and expressible in a form that collaborators without technical backgrounds can evaluate and that other researchers can reproduce.

The Mix-N-Map platform implements this transfer concretely \cite{kantan2024mixing}. In place of a code editor or scripting environment, it provides a structured visual interface in which movement signals from wearable sensors can be assigned to highly modular digital signal processing chains and connected to sound synthesis parameters in a variety of mapping configurations (ranging from one-to-one to many-to-many correspondences), all in real time on live or recorded movement data. Every parameter is modifiable while the sound is running: a designer adjusts how a particular movement dimension is handled and immediately hears the effect, without any programming or offline recompilation. The state of the interface at any moment constitutes a complete, reproducible record of the current design, serving as a snapshot that can be saved, shared with collaborators, or used as the basis for a written description. In clinical development contexts, this means a physiotherapist and a researcher can watch a patient's movement being sonified together, make adjustments in response to what they hear and observe, and save the resulting configuration for the next session, all within the flow of a clinical visit. This converts what would otherwise be a slow sequential development cycle (modify code, test offline, evaluate, repeat) into the kind of rapid iterative process that the FASTER framework \cite{wang2020faster} demands.

\section{A User-Centered Framework for Therapeutic Sonic Interaction}

\begin{figure}[H]
    \centering
    \includegraphics[width=0.9\textwidth]{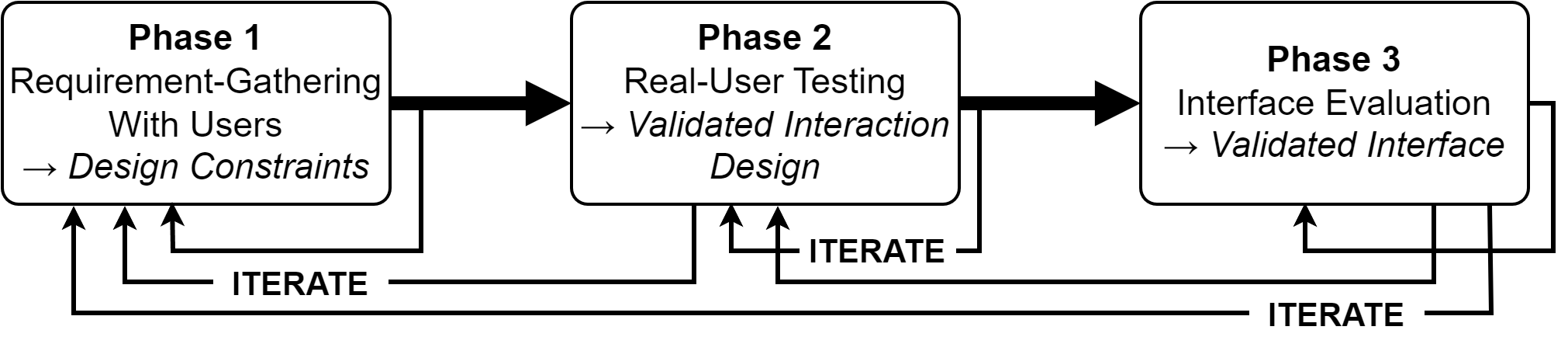}
    \caption{The three-phase user-centered design framework for therapeutic sonic
    interaction. Requirement-gathering with users in Phase~1 yield non-negotiable design
    constraints; real-user testing in Phase~2 produces a validated interaction
    design; and interface evaluation in Phase~3 yields a validated practitioner
    interface ready for deployment. All combinations of within/between phase iterations are possible depending on phase outcomes.}
    \label{fig:ucd_fw}
\end{figure}

The conceptual reframing of the signal-noise problem and the mixing-inspired design platform together address what the design process should focus on and what type of tools it should use. The remaining question is how stakeholder input should be structured and integrated throughout. The framework described here draws from our past work \cite{kantan2023orchestrating,kantan2024water} and positions user-centered design as the process through which design requirements and constraints are established in the first place: the clinical, perceptual, and practical requirements that define good outcomes cannot be fully anticipated in advance, and the gap between anticipated and actual requirements makes stakeholder engagement indispensable. The framework comprises three iterative phases as shown in Fig. \ref{fig:ucd_fw}. 

\begin{enumerate}
\item \textbf{Phase 1:} Here, requirements-gathering activities with practitioners and, where the domain demands it, members of the target user population establish the functional requirements that any workable system must satisfy before implementation begins. These activities may take several forms — structured practitioner interviews, observational studies, baseline data collection from the target population, or some combination — but their shared purpose is to ground subsequent design decisions in empirical knowledge of the clinical context and the people within it. The requirements they produce cover aspects such as setup time, interface vocabulary, what adjustments must be possible within the physical flow of a session, and what phenomena the feedback must be calibrated against \cite{kantan2024water}. Eliciting these explicitly before prototyping begins prevents investment in solutions that are technically impressive but clinically unworkable, and the necessity of doing so is constant across all application domains and care settings \cite{chandran2020exploring,or2022human}.
\item \textbf{Phase 2: } This concerns the iterative development and validation of the sonic interaction itself: which aspects of the person's behavior, state, or environment are reflected in the sound, how the connection from measurement to sound output is configured, and whether the resulting sound communicates the intended information under real conditions of use. Although this phase initially does involve practitioners standing in as patient proxies to understand the workings and capabilities of the system, the aforementioned aspects cannot be evaluated by the design team alone or by proxy testing; it requires testing with actual users in real settings, because the full range of variability in the target population (in needs, physical capacity, attentional load, and engagement with the task) is not reliably approximated by any substitute. The practitioner's role in these sessions is to evaluate whether the interaction addresses each user's therapeutic goals, grounding the assessment of perceptual function in clinical judgment. The outcome is a validated interaction paradigm with defined parameters and evidence of both perceptual legibility and clinical relevance for the target population. The need to test the interaction with real users (e.g. patients) while involving the clinical practitioner in judging therapeutic relevance applies regardless of application domain, whether the intervention involves movement, breathing, attention, or any other dimension of experience that sound can reflect \cite{stanton2017biofeedback,sigrist2013augmented,giggins2013biofeedback}.
\item \textbf{Phase 3:} This phase shifts focus from the interaction paradigm to the interface through which it is delivered: the system a practitioner uses to configure, monitor, and adjust the ABF technology during a session. Where Phase 2 asks whether the sound works for the person receiving it, Phase 3 asks whether the practitioner can operate the system independently within the real constraints of their working context. Development proceeds through practitioner interviews that establish non-negotiable interface requirements before implementation begins, followed by think-aloud sessions in which practitioners attempt representative tasks while verbalizing their thought process, exposing mismatches between design intent and actual use in real time. The culminating step is a feasibility study in a real deployment context, which reveals failure modes invisible to any prior testing. The latter may include the physical and cognitive demands of managing technology while simultaneously attending to a person in their care \cite{hamilton2019experiences,hamilton2022usability,wang2020faster}. The outputs are the primary evidence base for the next development iteration, and the principle that interface requirements must be derived from and validated against the practitioner's actual working conditions, rather than assumed from the designer's model of those conditions, holds across all application domains and care settings \cite{chandran2020exploring,or2022human}.
\end{enumerate}


\section{Case Study: HearWalk}

\begin{figure}[H]
    \centering
    \includegraphics[width=\textwidth]{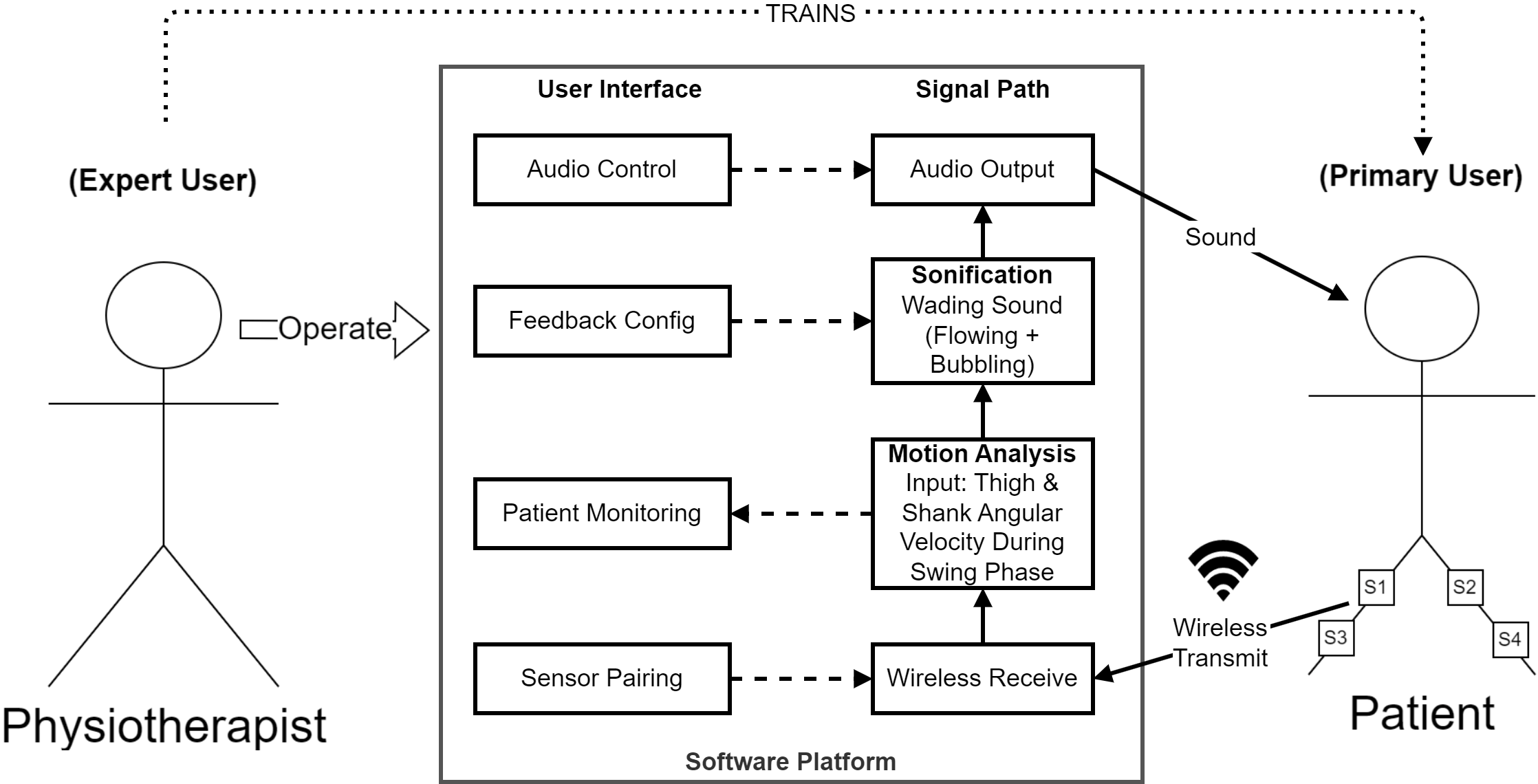}
    \caption{Overview of the HearWalk system architecture and signal path. The physiotherapist
    (expert user) operates the software platform to set up sensor connections, configure the
    feedback, monitor training in real time, and control the audio output. Wearable inertial sensors mounted on the patient's
    thighs and shanks (S1--S4) transmit angular velocity data wirelessly to the platform, which
    analyzes movement during the swing phase and drives a real-time sonification engine producing
    wading sounds whose energetic properties reflect instantaneous limb swing velocity. The
    resulting audio is returned to the patient, providing continuous feedback on gait symmetry and
    vigor.}
    \label{fig:system_combined}
\end{figure}

The HearWalk project serves to illustrate the three-phase framework applied in sequence, from initial requirements elicitation and algorithm development through clinical feasibility evaluation of a complete deployable system (see Fig. \ref{fig:system_combined}). The overarching goal was to design and build an ABF system that physiotherapists in the Danish public health system could operate independently during gait training with hemiparetic stroke patients, without requiring any technical expertise. The sonic paradigm at the heart of the system is an ecologically grounded design in which the angular velocities of the patient's thighs and shanks during walking are
mapped to the amplitude and spectral properties of water wading sounds. The rationale of this design rests on the intuitive association between the vigor of leg movement and the energetic quality of wading through shallow water \cite{kantan2023ecological,kantan2024wading}. The development of this system proceeded through all three framework phases, with each phase building directly on what the preceding one established.

\textbf{Phase 1 (Requirements Elicitation):} Before any algorithm implementation began, two parallel requirements-gathering activities anchored the project in clinical reality. The first was a kinematic baseline study: inertial sensor data were collected from fifteen hemiparetic patients during overground walking, yielding recordings of thigh and shank angular velocity trajectories
whose shapes captured the specific patterns of impairment (reduced range of motion, compensatory strategies, bilateral gait asymmetry) that any feedback algorithm would need to be designed against \cite{kantan2023ecological}. Graphical analysis of these trajectories, cross-referenced with concurrent video footage reviewed by an expert physiotherapist, confirmed that angular
velocity signals faithfully reflected clinically meaningful gait differences across patients with widely differing levels of functional ability. This representative kinematic dataset established the empirical ground truth for subsequent algorithm development, in that designs were calibrated against data from the actual target population rather than healthy-user approximations. The second requirements-gathering activity was the systematic definition of feedback design criteria drawn from the clinical and motor learning literature. Specifically, the feedback had to operate concurrently and without noticeable latency, use wireless wearable sensors, remain compatible with existing overground gait training protocols, accommodate the inherent variability of impaired movement through patient-specific configurability, and ground the movement-to-sound translation in an
intuitive embodied metaphor rather than an abstract acoustic encoding \cite{kantan2023ecological}. These criteria defined the boundaries of the design space before any prototype existed.

\textbf{Phase 2 (Algorithm Development and Clinical Validation):} With kinematic data in hand and requirements defined, three candidate feedback algorithms were developed and subjected to a structured expert evaluation, each employing a different sound model that converted limb swing angular velocity to variations in musical articulation dynamics: 
\begin{itemize}
\item Ecological wading simulation with movement-driven water sounds with optional movement-independent ambient background music
\item Harmonically rich abstract waveform with movement-driven melodic pitch changes between subsequent steps
\item Physics-based violin simulation with movement-driven melodic pitch changes between subsequent steps \cite{kantan2023ecological} 
\end{itemize}
A hands-on testing session with a focus group of five specialist neurorehabilitation physiotherapists assessed all three. The physics-based and abstract waveform algorithms were rejected on perceptual and clinical grounds as their artificial timbres were judged intolerable for sustained clinical use, and the pitch differences between subsequent steps perceptually interfered with the movement-driven amplitude variations that represented the movement trajectories themselves. The ecological wading algorithm was retained as the sole candidate, with the focus group citing its natural velocity-energy correspondence as both intuitive and clinically credible. Two design concerns were also raised. The algorithm's optional ambient music layer was perceived as conducive to relaxation rather than active movement, and the plain wading approach lacked a clear sonic goal that patients could work toward. Both concerns shaped the
subsequent redesign \cite{kantan2023ecological}.

The revised algorithm improved the quality of the core wading texture and added two goal-oriented variants in response to the physiotherapists' request for clearer sonic targets: 
\begin{itemize}
\item Positive reinforcement variant that triggered an impulsive splash sound when the shank angular velocity crossed a configurable threshold. This gave the patient a discrete sonic reward to aim for on each step
\item Negative reinforcement variant in which a contextually aversive sound (urination) dominated at low movement velocities and was progressively masked by the louder wading sounds as velocity increased, motivating the patient to swing more forcefully to
suppress it \cite{kantan2023ecological}.
\end{itemize}
A feasibility study with nine hemiparetic patients and seven physiotherapists tested all three variants (plain, positive reinforcement, negative reinforcement) during real overground gait training sessions in a rehabilitation hospital setting. The results were broadly validating of the ecological wading approach. The majority of patients found the wading sounds immediately natural
and relatable. Several explicitly likened the experience to walking through water without any instruction to interpret the sound and reported a sharpened awareness of gait rhythm and the asymmetry between their two sides. Kinematic analysis of sensor recordings corroborated therapist-observed improvements in cadence consistency and bilateral symmetry for a subset of patients during the feedback condition, with one patient's therapist observing that the feedback-assisted session showed the best walking quality they had witnessed from that patient to date \cite{kantan2023ecological}. Patients who reported using the sound
actively tended to describe spontaneous strategies such as attempting to equalize the sound between the two sides, or timing their steps to produce a rhythmic pattern in the wading sounds. The patients were readily able to understand the movement-sound relationship and follow simple instructions such as `push more water with your left side', suggesting that the ecological metaphor successfully enabled patients to translate auditory information into movement adjustments without a deliberate decoding step. 

One of the two goal-oriented variants produced more mixed findings that connect directly to the signal-noise paradox introduced earlier. The negative reinforcement sound (a urination sound intended to dominate during low-velocity movement) shared substantial spectral content with the broadband wading texture it was embedded in. Because the wading sounds themselves carry considerable stochastic variation across the audible range, the urination signal had no perceptually clear foreground to emerge from during the low-velocity conditions where it mattered most. Several therapists reported difficulty detecting when a patient was moving poorly, with two explicitly citing the urination variant as the problem case \cite{kantan2023ecological}. This is precisely the failure mode the noise profile concept describes: the intrinsic variation of the chosen sound model masked the movement-driven variation carrying the clinically meaningful information because the two sound models did not leave adequate spectral room for the signal to be distinguishable. Two therapists also noted that layering multiple sounds imposed a cognitive demand on patients simultaneously managing an effortful motor task, and recommended starting with a single focused sound before introducing additional layers — which is, in signal-noise terms, a reduction in model-intrinsic variability that restores the conditions under which movement-driven changes can be perceived \cite{kantan2023ecological}. 

\textbf{Phase 3 (Interface Development and Clinical Feasibility):} The third phase addressed the transition from a validated feedback algorithm to a system designed for clinical deployment. An initial structured interview with three physiotherapists established the non-negotiable interface requirements before any implementation began \cite{kantan2024making}: 
\begin{itemize}
\item Setup including sensor mounting should be completable in approximately fifteen minutes.
\item The interface must use terminology familiar to clinicians rather than engineering or signal-processing language.
\item Feedback sensitivity (responsiveness to movement) must be adjustable one-handed while simultaneously supporting the patient.
\item Automatic calibration to each patient's individual gait parameters should reduce the manual adjustment burden to a minimum \cite{kantan2024water}.
\end{itemize}
These requirements directly shaped both the system architecture and the structure of the interface. 

Over a six-month development period, three formative evaluation sessions with naive physiotherapists using think-aloud protocols refined the screen layout, corrected terminology mismatches, and clarified the sequential workflow a therapist would follow from session start to finish. A focus group session with five physiotherapists, conducted before any patient testing, surfaced further
issues with control labelling and screen organization and informed a near-final interface revision \cite{kantan2024water}. The resulting interface was organized as a linear sequence of five screens that mirrored the natural procedural structure of a clinical gait training visit: main menu, sensor connection, sensor placement, training, and session summary. Proceeding through these screens in order would take a physiotherapist from system startup to an active feedback session without requiring any decision-making outside of the clinical workflow itself. The training screen, which was the primary operational view during a session, provided real-time visual monitoring of limb angular velocities, a feedback sensitivity slider, and controls for adjusting the feedback to one or both sides of the body. An automatic calibration function updated the angular velocity ranges based on ongoing movement data, adapting the feedback to each patient's kinematics without requiring manual parameter entry.

\begin{figure}[H]
    \centering
    \includegraphics[width=\textwidth]{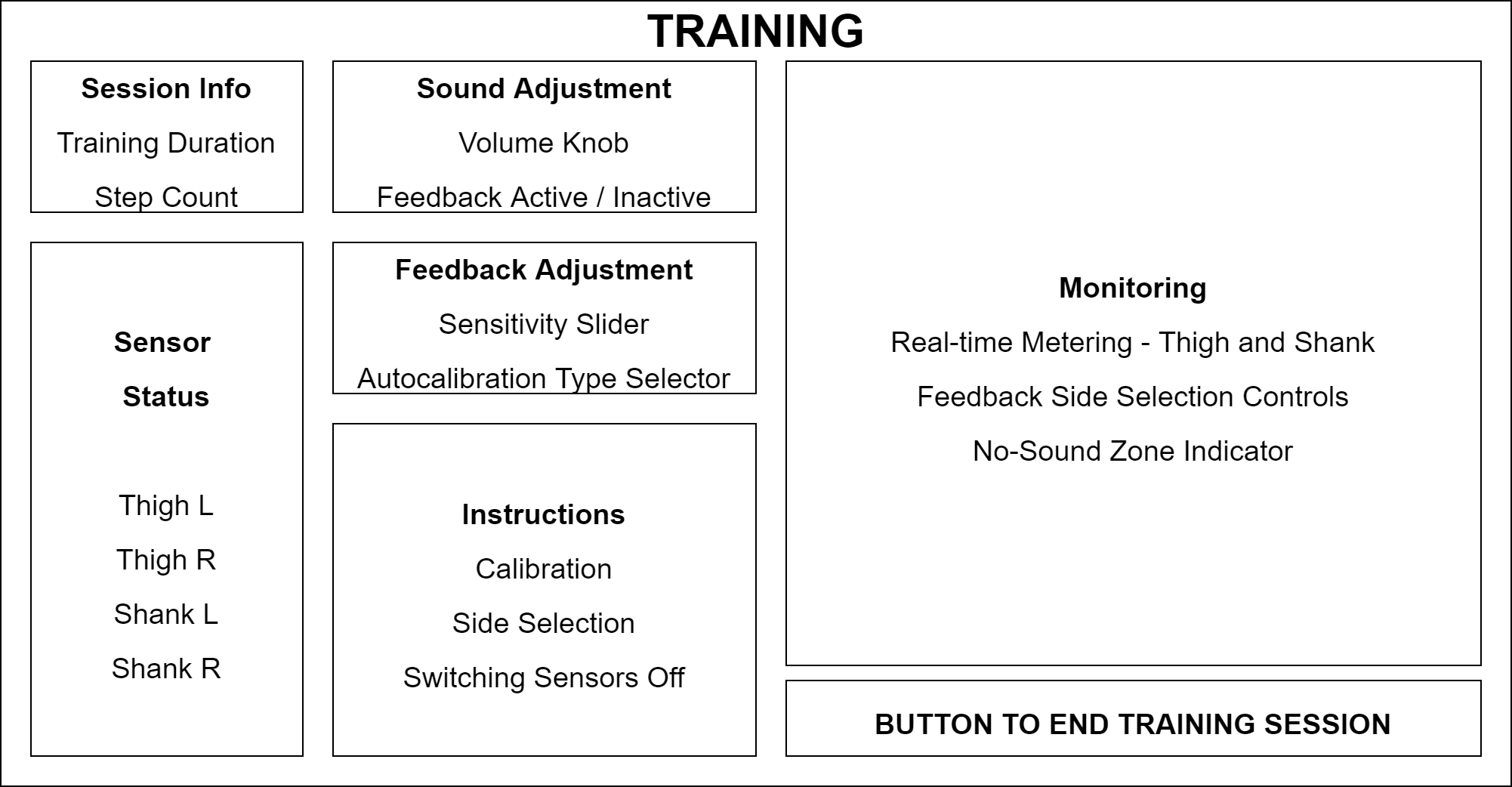}
   \caption{Schematic of the HearWalk training screen. Reading left to right, the left panel displays session info and sensor status, the central panel groups all feedback configuration controls, and the right panel provides real-time angular velocity monitoring across the four sensor locations. Therapists in the feasibility study worked with a Danish-language version of this interface.}
\label{fig:hearwalk}
\end{figure}

The interface was implemented in Danish to match the primary language of the clinical setting. As shown in Fig.~\ref{fig:hearwalk}, the training screen presents four angular velocity meters (one per sensor location: left and right thigh, left and right shank), a step counter, a session timer, and a central sensitivity slider labelled \textit{Sensitivitet} ranging from \textit{Lav}
(low) to \textit{H{\o}j} (high). At low sensitivity, only large, vigorous limb swings produce clearly audible wading sounds, making asymmetry maximally salient for patients with severe gait impairment; at high sensitivity, even modest limb movement generates a clear acoustic response, suitable for patients with relatively preserved gait who need fine-grained feedback on subtle
kinematic differences. The physiotherapist could adjust this slider on the fly based on what they observed in the patient's movement and heard in the feedback---a design that placed the clinical judgment about what the patient needs in the hands of the practitioner rather than embedding it in an algorithm. The language of the interface elements was intentionally drawn from clinical practice, with controls describing what the physiotherapist wanted the system to do for the patient, not what the underlying signal processing was computing.

The final prototype was evaluated in a feasibility study with four physiotherapists and seven hemiparetic patients in a rehabilitation hospital setting. All four physiotherapists found setup straightforward and reported the workflow as naturally compatible with their existing training routines; the automatic calibration function drew no complaints of added workload. The ecological
wading sounds were found by several patients to be immediately natural and understandable. Kinematic improvements in cadence and gait symmetry were visible for one patient over the course of the session. The evaluation also uncovered the gaps that would define the next potential development iteration:
\begin{itemize}
\item The laptop-based system placed on a corner table prevented effective use of monitoring and adjustment functions by the therapist when patients required hands-on physical support.
\item The single available feedback sound aimed at normal-pace overground walking was insufficient for the full range of patient presentations and training goals that the study encountered.
\item Physiotherapists reported needing additional practice before being able to use the sensitivity controls fluently in
parallel with conducting a training session \cite{kantan2024water}.
\end{itemize}
These findings were the type of expected output of a well-executed Phase 3 feasibility evaluation, which was designed precisely to uncover the practical limitations that must be addressed before scaled deployment is meaningful. Taken together, the three phases of the HearWalk project demonstrate the framework as a connected sequence rather than a set of independent activities. Phase 1 established the kinematic and clinical grounding for algorithm design before any implementation began. Phase 2 drove the iterative convergence on the ecological wading paradigm through direct exposure with patients and physiotherapists in real clinical conditions, discarding candidates that failed on perceptual or clinical grounds and validating the core design concept through observed kinematic change and patient-reported experience. Phase 3 translated that validated paradigm into a system designed around the operational constraints of clinical practice, uncovering implementation challenges invisible to any earlier stage of testing. Each phase produced not only a design output but a specification of the requirements for the next phase, making the development trajectory explicit and evidence-grounded throughout. Despite the known costs of clinical user-centered processes in terms of time, access, and ethical preparation \cite{schnall2016user}, our experience with HearWalk adds to a growing body of evidence that each phase of testing is a necessity, as real clinical testing conditions routinely reveal failure modes that no proxy testing can reproduce, and it is precisely those types of findings that help make subsequent development iterations tractable and specifically actionable.

\section{AI-Augmented Sonification Design: Possibilities and Limits}

The framework documented in the preceding sections rests on two kinds of specialized work that are jointly necessary but in practice often constrained: the iterative empirical process of navigating a multidimensional sonification design space, and the structured clinical engagement that establishes the design constraints that define good outcomes. The HearWalk case study illustrates both the value of this combination and its costs, i.e., a six-month development cycle, multiple rounds of formative evaluation, and a clinical feasibility study that revealed further gaps requiring subsequent iteration. These costs reflect structural features of the field such as the cross-disciplinary demands of clinical sonification development, and the irreducible complexity of deploying technology in therapeutic contexts. It is against this backdrop that the emergence of capable \emph{large language models (LLMs)} and multimodal AI systems becomes relevant. These tools are potentially useful across the entire pipeline from early ideation to deployable clinical applications. We next explore their potential contributions and limitations at each stage in turn.

\paragraph{Ideation.} At the earliest stages of design, LLMs can function as generative interlocutors. In other words, given a description of the clinical \emph{context}, the movement \emph{phenomenon} to be sonified, and the intended therapeutic \emph{goal}, they can propose candidate mapping strategies, articulate rationales for particular sound-parameter correspondences, and extrapolate ideas from other sonification applications \cite{zhou2025llmdesign}. Because they can produce large numbers of candidate concepts in a short time, they can help designers explore a wider design space than would be practical through manual literature search and individual deliberation alone \cite{li2025llmideation}. There is broader evidence from creative and innovation contexts that AI-generated options can expand the range of solutions a human designer considers, even when those options are not themselves adopted directly \cite{doshi2024generative,haase2024human}. In the HearWalk context, for instance, LLM-assisted ideation at the outset of the project might have surfaced candidate sonic metaphors that we could then have evaluated against the ecological grounding criteria that ultimately motivated the water wading paradigm. At the same time, LLMs have no embodied experience of sound perception, no access to clinical knowledge about what specific patient populations can attend to under movement task demands, and no principled grounding in the perceptual science of auditory-motor coupling \cite{xu2025llmgrounding}. Their suggestions are pattern-matched from training data and may reproduce plausible-sounding but poorly calibrated assumptions about, for example, which auditory dimensions most reliably convey spatial or temporal movement information to patients with neurological impairment. Ideation assistance therefore requires expert filtering in the form of a clinical researcher or sonification specialist who can identify which AI-generated concepts merit development and which rest on false assumptions about the target context.

\paragraph{Sketching.} Once a candidate concept has been identified, AI tools can assist in translating an informal verbal description into a more structured specification of which movement parameters map to which acoustic dimensions, with what scaling functions, and with what intended perceptual effect. LLMs can generate pseudocode, parameter tables, or structured mapping descriptions from conversational input. This can help designers articulate their intentions more precisely and reveal ambiguities or underspecified aspects of a design concept before any implementation effort is invested \cite{li2025llmideation}. Multimodal models capable of processing movement data or sensor traces alongside text could, in principle, suggest mapping functions calibrated to the statistical properties of the signal, for instance recommending a logarithmic rather than linear scaling function given a movement variable with a skewed distribution. The limitation at this stage is that a sketch shaped by AI inherits whatever distributional biases and domain gaps the model carries. Without explicit grounding in the perceptual requirements of the clinical scenario (patient hearing profile, attentional capacity, background noise environment), AI-generated sketches may be internally coherent but not fit for a given purpose. In the HearWalk design process, translating the informal concept of `water wading sounds reflecting gait energy' into a precise algorithmic specification required iterative expert judgment. Current AI sketching tools could potentially carry out such concept-to-algorithm translations with far greater speed if provided with detailed and clinically informed prompting. Human expert review of the sketch against clinical and perceptual criteria remains necessary before any implementation is begun.

\paragraph{Prototyping.} This is the stage at which AI assistance is currently most practically developed. LLMs can generate functional audio mapping and synthesis code from natural language design descriptions, substantially reducing the programming barrier (both in terms of ability and speed) that has historically limited who can participate in sonification design \cite{jiang2025codegen}. 
LLMs can thus effectively compress the time between design concept and auditable prototype, directly accelerating the iterative UCD cycle described in the previous section.
However, there are substantial limitations at this stage that are worth noting. LLMs generate code by pattern-matching against training data, and in domains where relevant patterns are underrepresented (e.g. specialized real-time audio or sonification libraries) they hallucinate, producing functionally plausible-looking code that references non-existent methods or makes incorrect assumptions about the behavior of application programming interfaces \cite{liu2024exploring,tian2025codehalu}. Studies of LLM code generation across several languages and tasks at the time of writing have found that 5 to 22\% of generated outputs contain hallucinated packages or API calls \cite{spracklen2025we}, and the perverse property of such errors is that they often fail in ways that are not immediately visible, passing superficial inspection and manifesting only under specific runtime conditions. In a clinical context, where the code controls a sound system that patients and physiotherapists are relying on during training, a subtle bug in the audio processing pipeline could produce feedback that sounds plausible but systematically misrepresents movement in a manner potentially counterproductive to motor learning \cite{moinuddin2021feedback}. Beyond code reliability, current LLMs lack grounding in embodied perception, in that they can produce a syntactically correct mapping function, but have no capacity to evaluate whether the resulting sound will actually convey the intended kinematic information to a patient under the attentional demands of rehabilitation \cite{xu2025llmgrounding}. Effective prototyping with AI therefore requires human oversight at multiple checkpoints: context engineering to articulate design goals and constraints clearly, validation of generated code against behavioral requirements, perceptual evaluation of the sonification against fidelity and aesthetic criteria, and clinical appropriateness checking against the target population's cognitive load profile.

\paragraph{Developing deployable applications.} Moving from prototype to a robust, clinically deployable application involves tasks where AI assistance is potentially useful but where the stakes of failure are highest. AI tools can assist with generating documentation, producing configuration interfaces, writing test scaffolding, and refactoring prototype code into maintainable structures \cite{jiang2025codegen}. In the HearWalk project, transitioning the validated prototype to a clinically deployable system required generating physiotherapist-oriented documentation, producing configuration guides in clinical rather than engineering terminology, and adapting the interface to the operational workflows of the Danish public health system. In all these tasks, AI text-generation and documentation tooling could have meaningfully reduced the team's overhead, even if the clinical feasibility evaluation would still remain necessary regardless. Automated code review and static analysis tools augmented with AI can flag common security vulnerabilities, accessibility shortfalls, and performance bottlenecks before human review \cite{li2024iris}. At the same time, clinical software is subject to regulatory requirements (e.g. medical device legislation in many jurisdictions) that impose validation, traceability, and post-market surveillance obligations that AI tools are not equipped to satisfy autonomously \cite{mesko2023llmregulation}. Any AI involvement in the development of regulated software may need to be explicitly documented and justified in regulatory submissions, and AI-generated code reaching this stage must undergo thorough audit by qualified developers \cite{ebinger2025fdareg}. There is also a longer-term maintenance consideration: AI-generated codebases may accumulate technical debt if the generated code lacks the modular organization and documentation conventions that allow human developers to extend and debug it efficiently over time \cite{moreschini2025techdebt,bogner2021characterizing}.

\paragraph{Overall positioning.} Across all stages, we position AI as a design assistant rather than an autonomous designer \cite{song2024human}, with a role that varies in character across the pipeline. It may be generative and exploratory at the ideation stage, structuring and formalizing at the sketching stage, productive but requiring close validation at the prototyping stage, and supportive but subordinate to regulatory and clinical quality standards at the deployment stage. This human-in-the-loop architecture is the mechanism that makes AI participation compatible with the quality standards that therapeutic applications demand. Looking forward, more ambitious integrations become conceivable as models gain better grounding in perceptual and clinical domains. 
Realizing the potential of LLM integration requires the development of validation frameworks specific to generative systems in clinical contexts \cite{alelyani2025aiframework}, clear explainability standards for AI-assisted design decisions \cite{mienye2024xai}, and professional training that prepares researchers and clinicians to supervise rather than defer to AI-assisted design processes.

\section{Conclusions}

This chapter has presented a framework for designing therapeutic sonic interactions built around three interconnected contributions: a conceptual reframing of the design task as the calibration of sonic variability to the perceptual affordances of the listener and the clinical context; a mixing-inspired platform that imposes a structured, real-time workflow on the exploration of sonic interaction designs; and a development framework that grounds design decisions in iterative engagement with clinicians and patients. The HearWalk case study traces a full development arc across all three framework phases, from kinematic data collection and requirements elicitation, through iterative algorithm design and clinical validation of the ecological wading paradigm, to the design and feasibility evaluation of a complete physiotherapist-operated interface. At each stage, testing with real users in real clinical conditions uncovered requirements and failure modes that no earlier stage of the process could have anticipated, and it was precisely those findings that guided each subsequent iteration. Several directions remain open. The framework has been developed primarily in the context of post-stroke gait rehabilitation, and extending it to balance training, upper limb training, pain management, and other applications requires both generalizability testing of the existing design guidelines and investigation of potential context-specific requirements not yet encountered. Within the FASTER development process, the work presented here covers the development phase in full; scaled evaluation studies and implementation research constitute the necessary next steps toward a clinical tool with reliable evidence behind it \cite{wang2020faster}. The role of AI tools, examined in the preceding section, follows a stage-differentiated logic in which the human-in-the-loop architecture is the enabling condition for AI participation at every stage. The ultimate aim is to make the auditory channel a practical and reliable medium for supporting therapeutic goals through principled design and engineering practices centered around the needs and abilities of users.

\section{Acknowledgments}

We would like to thank Head Research Therapist Helle Rovsing M. Jørgensen, the physiotherapists at Neuroenhed Nord (North Denmark Regional Hospital) and the patients who participated in the HearWalk project. The project was funded by BETA.HEALTH (The Danish National Innovation Platform for Future Healthcare) grant number 2022-1294. The contribution of P.R.K. and S.D. to HearWalk was partially funded by NordForsk’s Nordic University Hub, Nordic Sound and Music Computing Network NordicSMC, project number 86892.

\bibliographystyle{splncs04}
\bibliography{mybibliography}

@book{thaut2014nmtBook,
  address={New York, NY, US},
  title={Handbook of Neurologic Music Therapy},
  isbn={978-0-19-969546-1},
  publisher={Oxford University Press},
  editor={Thaut, Michael H and Hoemberg, Volker},
  year={2014}
}

@phdthesis{kantan2023orchestrating,
  title={Orchestrating Motion: Real-Time Sonification Tools and Methods to Support Movement Rehabilitation},
  author={Kantan, Prithvi Ravi},
  year={2023},
  school={Aalborg University},
  doi={10.54337/aau696001045}
}

@article{kantan2024water,
  title={Making Movement Sonification Usable in Clinical Gait Rehabilitation: A User-Centered Study},
  author={Kantan, Prithvi Ravi and Dahl, Sofia and J{\o}rgensen, Helle Rovsing and Spaich, Erika G},
  journal={Audio Mostly 2024 -- Explorations in Sonic Cultures},
  year={2024},
  publisher={ACM},
  doi={10.1145/3678299.3678304}
}

@inproceedings{bogner2021characterizing,
  title={Characterizing technical debt and antipatterns in AI-based systems: A systematic mapping study},
  author={Bogner, Justus and Verdecchia, Roberto and Gerostathopoulos, Ilias},
  booktitle={2021 IEEE/ACM International Conference on Technical Debt (TechDebt)},
  pages={64--73},
  year={2021},
  organization={IEEE}
}

@article{kantan2024wading,
  title={Sonifying gait kinematics: generating water wading sounds through a digital Foley approach},
  author={Kantan, Prithvi Ravi and Dahl, Sofia and Serafin, Stefania and Spaich, Erika G},
  journal={Personal and Ubiquitous Computing},
  year={2024},
  doi={10.1007/s00779-024-01829-1}
}

@article{kantan2023ecological,
  title={Designing Ecological Auditory Feedback on Lower Limb Kinematics for Hemiparetic Gait Training},
  author={Kantan, Prithvi Ravi and Dahl, Sofia and J{\o}rgensen, Helle Rovsing M{\o}ller and Khadye, Chetali and Spaich, Erika G},
  journal={Sensors},
  volume={23},
  number={8},
  pages={3964},
  year={2023},
  doi={10.3390/s23083964}
}

@inproceedings{kantan2024mixing,
  title={Adapting Audio Mixing Principles and Tools to Parameter Mapping Sonification Design},
  author={Kantan, Prithvi Ravi and Dahl, Sofia and Spaich, Erika G},
  booktitle={Proceedings of the 29th International Conference on Auditory Display (ICAD 2024)},
  pages={148--155},
  year={2024},
  doi={10.21785/icad2024.034}
}

@inbook{kos2018bmbf,
  author={Kos, Anton and Umek, Anton},
  year={2018},
  title={Biomechanical Biofeedback},
  isbn={978-3-319-91348-3},
  publisher={Springer},
  doi={10.1007/978-3-319-91349-0_2}
}

@article{sigrist2013augmented,
  title={Augmented Visual, Auditory, Haptic, and Multimodal Feedback in Motor Learning: A Review},
  author={Sigrist, Roland and Rauter, Georg and Riener, Robert and Wolf, Peter},
  journal={Psychonomic Bulletin \& Review},
  volume={20},
  number={1},
  pages={21--53},
  year={2013},
  publisher={Springer},
  doi={10.3758/s13423-012-0333-8}
}

@article{guerra2020soniPhysio,
  title={The Use of Sonification for Physiotherapy in Human Movement Tasks: A Scoping Review},
  author={Guerra, Joao and Smith, Lee and Vicinanza, Domenico and Stubbs, Brendon and Veronese, Nicola and Williams, G},
  journal={Science \& Sports},
  volume={35},
  number={3},
  pages={119--129},
  year={2020},
  publisher={Elsevier}
}

@article{stanton2017biofeedback,
  title={Biofeedback improves performance in lower limb activities more than usual therapy in people following stroke: a systematic review},
  author={Stanton, Rosalyn and Ada, Louise and Dean, Catherine M and Preston, Elisabeth},
  journal={Journal of Physiotherapy},
  volume={63},
  number={1},
  pages={11--16},
  year={2017},
  publisher={Elsevier}
}

@article{giggins2013biofeedback,
  author={Giggins, Oonagh and McCarthy Persson, Ulrik and Caulfield, Brian},
  year={2013},
  title={Biofeedback in Rehabilitation},
  volume={10},
  pages={60},
  journal={Journal of Neuroengineering and Rehabilitation},
  doi={10.1186/1743-0003-10-60}
}

@article{ma2016balance,
  author={Ma, Christina and Wong, Duo and Lam, Gilbert W K and Wan, Anson and Lee, Winson},
  year={2016},
  title={Balance Improvement Effects of Biofeedback Systems with State-of-the-Art Wearable Sensors: A Systematic Review},
  volume={16},
  pages={434},
  journal={Sensors},
  doi={10.3390/s16040434}
}

@article{lahav2007action,
  title={Action representation of sound: audiomotor recognition network while listening to newly acquired actions},
  author={Lahav, Amir and Saltzman, Elliot and Schlaug, Gottfried},
  journal={Journal of Neuroscience},
  volume={27},
  number={2},
  pages={308--314},
  year={2007},
  publisher={Soc Neuroscience}
}

@article{bevilacqua2016sensorimotor,
  title={Sensori-motor learning with movement sonification: perspectives from recent interdisciplinary studies},
  author={Bevilacqua, Fr{\'e}d{\'e}ric and Boyer, Eric O and Fran{\c{c}}oise, Jules and Houix, Olivier and Susini, Patrick and Roby-Brami, Agn{\`e}s and Hanneton, Sylvain},
  journal={Frontiers in Neuroscience},
  volume={10},
  pages={195146},
  year={2016},
  publisher={Frontiers}
}

@article{kramer2010sonification,
  title={Sonification Report: Status of the Field and Research Agenda},
  author={Kramer, Gregory and Walker, Bruce and Bonebright, Terri and Cook, Perry and Flowers, John H and Miner, Nadine and Neuhoff, John},
  year={2010}
}

@article{grond2011parameter,
  title={Parameter mapping sonification},
  author={Grond, Florian and Berger, Jonathan},
  journal={The Sonification Handbook},
  pages={363--397},
  year={2011},
  publisher={Logos Verlag Berlin}
}

@inproceedings{hermann2008taxonomy,
  title={Taxonomy and definitions for sonification and auditory display},
  author={Hermann, Thomas and Hunt, Andy and Neuhoff, John G},
  booktitle={Proceedings of the 14th International Conference on Auditory Display},
  year={2008}
}

@article{schaffert2019review,
  title={A review on the relationship between sound and movement in sports and rehabilitation},
  author={Schaffert, Nina and Braun Janzen, Thenille and Mattes, Klaus and Thaut, Michael H},
  journal={Frontiers in Psychology},
  volume={10},
  pages={244},
  year={2019},
  publisher={Frontiers Media SA}
}

@article{effenberg2005movement,
  title={Movement sonification: Effects on perception and action},
  author={Effenberg, Alfred O},
  journal={IEEE Multimedia},
  volume={12},
  number={2},
  pages={53--59},
  year={2005},
  publisher={IEEE}
}

@article{dyer2016sonification,
  title={Sonification of movement for motor skill learning in a novel bimanual task: Aesthetics and retention strategies},
  author={Dyer, John and Stapleton, Paul and Rodger, Matthew},
  booktitle={Proceedings of the 22nd International Conference on Auditory Display},
  pages={99--102},
  year={2016}
}

@article{dyer2015movementSoni,
  title={Sonification as Concurrent Augmented Feedback for Motor Skill Learning and the Importance of Mapping Design},
  author={Dyer, John F and Stapleton, Paul and Rodger, Matthew W M},
  journal={The Open Psychology Journal},
  volume={8},
  number={1},
  year={2015}
}

@article{roddy2020mapping,
  title={Mapping for meaning: the embodied sonification listening model and its implications for the mapping problem in sonic information design},
  author={Roddy, Stephen and Bridges, Brian},
  journal={Journal on Multimodal User Interfaces},
  volume={14},
  number={2},
  pages={143--151},
  year={2020},
  publisher={Springer}
}

@article{leman2018embodied,
  title={What is embodied music cognition?},
  author={Leman, Marc and Maes, Pieter-Jan and Nijs, Luc and Van Dyck, Edith},
  journal={Springer Handbook of Systematic Musicology},
  pages={747--760},
  year={2018},
  publisher={Springer}
}

@article{craig2008developing,
  title={Developing and evaluating complex interventions: the new Medical Research Council guidance},
  author={Craig, Peter and Dieppe, Paul and Macintyre, Sally and Michie, Susan and Nazareth, Irwin and Petticrew, Mark},
  journal={BMJ},
  volume={337},
  year={2008},
  publisher={British Medical Journal Publishing Group}
}

@article{bleijenberg2018mrc,
  title={Increasing value and reducing waste by optimizing the development of complex interventions: Enriching the development phase of the Medical Research Council (MRC) Framework},
  author={Bleijenberg, Nienke and de Man-van Ginkel, Janneke M and Trappenburg, Jaap C A and Ettema, Roelof G A and Sino, Carolien G and Heim, Noor and Hafsteind{\'o}ttir, Th{\'o}ra B and Richards, David A and Schuurmans, Marieke J},
  journal={International Journal of Nursing Studies},
  volume={79},
  pages={86--93},
  year={2018},
  publisher={Elsevier}
}

@article{wang2020faster,
  title={The time is now: a FASTER approach to generate research evidence for technology-based interventions in the field of disability and rehabilitation},
  author={Wang, Rosalie H and Kenyon, Lisa K and McGilton, Katherine S and Miller, William C and Hovanec, Nina and Boger, Jennifer and Viswanathan, Pooja and Robillard, Julie M and Czarnuch, Stephen M},
  journal={Archives of Physical Medicine and Rehabilitation},
  volume={102},
  number={9},
  pages={1848--1859},
  year={2021},
  publisher={Elsevier}
}

@incollection{ross2015types,
  title={Types of intervention and their development},
  author={Ross, David A and Smith, Peter G and Morrow, Richard H},
  booktitle={Field Trials of Health Interventions, 3rd edition},
  year={2015},
  publisher={Oxford University Press}
}

@article{wade2005describing,
  title={Describing rehabilitation interventions},
  author={Wade, Derick T},
  journal={Clinical Rehabilitation},
  volume={19},
  number={8},
  pages={811--818},
  year={2005},
  publisher={SAGE Publications Sage CA: Thousand Oaks, CA}
}

@article{gould1985designing,
  title={Designing for usability: key principles and what designers think},
  author={Gould, John D and Lewis, Clayton},
  journal={Communications of the ACM},
  volume={28},
  number={3},
  pages={300--311},
  year={1985},
  publisher={ACM New York, NY, USA}
}

@incollection{or2022human,
  title={Human factors engineering and user-centered design for mobile health technology: enhancing effectiveness, efficiency, and satisfaction},
  author={Or, Calvin Kalun and Holden, Richard J and Valdez, Rupa S},
  booktitle={Human-Automation Interaction: Mobile Computing},
  pages={97--118},
  year={2022},
  publisher={Springer}
}

@inproceedings{chandran2020exploring,
  title={Exploring user centered design in healthcare: a literature review},
  author={Chandran, Srijith and Al-Sa'di, Ahmed and Ahmad, Esraa},
  booktitle={2020 4th International Symposium on Multidisciplinary Studies and Innovative Technologies (ISMSIT)},
  pages={1--8},
  year={2020},
  organization={IEEE}
}

@article{dabbs2009user,
  title={User-centered design and the development of patient decision aids: protocol for a systematic review},
  author={Dabbs, Annette DeVito and Myers, Bradley A and McLaughlin Crabtree, Mercy K and Thorn, Karen A and Stilley, Carol S and Hoffman, Judith and Sereika, Susan M},
  journal={Systematic Reviews},
  volume={4},
  year={2009}
}

@article{doshi2024generative,
  title={Generative AI enhances individual creativity but reduces the collective diversity of novel content},
  author={Doshi, Anil R and Hauser, Oliver P},
  journal={Science Advances},
  volume={10},
  number={28},
  pages={eadn5290},
  year={2024},
  publisher={American Association for the Advancement of Science}
}

@article{haase2024human,
  title={Human-AI co-creativity: Exploring synergies across levels of creative collaboration},
  author={Haase, Jennifer and Pokutta, Sebastian},
  journal={arXiv preprint arXiv:2411.12527},
  year={2024}
}

@article{liu2024exploring,
  title={Exploring and evaluating hallucinations in LLM-powered code generation},
  author={Liu, Fang and Liu, Yang and Shi, Lin and Huang, Houkun and Wang, Ruifeng and Yang, Zhen and Zhang, Li and Li, Zhongqi and Ma, Yuchi},
  journal={arXiv preprint arXiv:2404.00971},
  year={2024}
}

@inproceedings{tian2025codehalu,
  title={CodeHalu: Investigating Code Hallucinations in LLMs via Execution-Based Verification},
  author={Tian, Yuchen and Yan, Weixiang and Yang, Qian and Zhao, Xuandong and Chen, Qian and Wang, Wen and Luo, Ziyang and Ma, Lei and Song, Dawn},
  booktitle={Proceedings of the AAAI Conference on Artificial Intelligence},
  volume={39},
  number={24},
  pages={25300--25308},
  year={2025}
}

@inproceedings{spracklen2025we,
  title={We have a package for you! A comprehensive analysis of package hallucinations by code generating LLMs},
  author={Spracklen, Joseph and Wijewickrama, Raveen and Sakib, A H M Nazmus and Maiti, Anindya and Viswanath, Bimal},
  booktitle={34th USENIX Security Symposium (USENIX Security 25)},
  pages={3687--3706},
  year={2025}
}

@article{song2024human,
  title={Human-AI collaboration by design},
  author={Song, Binyang and Zhu, Qihao and Luo, Jianxi},
  journal={Proceedings of the Design Society},
  volume={4},
  pages={2247--2256},
  year={2024},
  publisher={Cambridge University Press}
}

@article{zhou2025llmdesign,
  author  = {Zhou, Yehong and Chen, Chun-Hsien},
  title   = {Examining the impact of large language models on design: {Functions}, strengths, limitations, and roles},
  journal = {Design and Artificial Intelligence},
  volume  = {1},
  number  = {2},
  pages   = {100017},
  year    = {2025},
  doi     = {10.1016/j.daai.2025.100017}
}

@misc{li2025llmideation,
  author        = {Li, Sitong and Padilla, Stefano and {Le Bras}, Pierre and Dong, Junyu and Chantler, Mike},
  title         = {A Review of {LLM}-Assisted Ideation},
  year          = {2025},
  howpublished  = {arXiv preprint arXiv:2503.00946},
  eprint        = {2503.00946},
  archivePrefix = {arXiv},
  primaryClass  = {cs.HC}
}

@article{bradt2021cochrane,
  author  = {Bradt, Joke and Dileo, Cheryl and Myers-Coffman, Katherine and Biondo, Jaclyn},
  title   = {Music interventions for improving psychological and physical outcomes in people with cancer},
  journal = {Cochrane Database of Systematic Reviews},
  volume  = {2021},
  number  = {10},
  pages   = {CD006911},
  year    = {2021},
  doi     = {10.1002/14651858.CD006911.pub4}
}

@article{koelsch2014brain,
  author  = {Koelsch, Stefan},
  title   = {Brain correlates of music-evoked emotions},
  journal = {Nature Reviews Neuroscience},
  volume  = {15},
  number  = {3},
  pages   = {170--180},
  year    = {2014},
  doi     = {10.1038/nrn3666}
}

@article{eddins1995temporal,
  title={Temporal integration and temporal resolution},
  author={Eddins, David A and Green, David M},
  journal={Hearing},
  pages={207--242},
  year={1995}
}

@article{demetriou2026scoping,
  author  = {Demetriou, Andrew M. and Bowling, Daniel L.},
  title   = {A scoping review of music-based digital therapeutics for stress, anxiety, and depression},
  journal = {Frontiers in Human Neuroscience},
  volume  = {20},
  pages   = {1602004},
  year    = {2026},
  doi     = {10.3389/fnhum.2026.1602004}
}

@article{nown2022mapping,
  author  = {Nown, Thomas H. and Upadhyay, Priti and Kerr, Andrew and Andonovic, Ivan and Tachtatzis, Christos and Grealy, Madeleine A.},
  title   = {A Mapping Review of Real-Time Movement Sonification Systems for Movement Rehabilitation},
  journal = {IEEE Reviews in Biomedical Engineering},
  volume  = {16},
  pages   = {672--686},
  year    = {2022},
  doi     = {10.1109/RBME.2022.3187840}
}

@article{xu2025llmgrounding,
  author  = {Xu, Qihui and Peng, Yingying and Nastase, Samuel A. and Chodorow, Martin and Wu, Minghua and Li, Ping},
  title   = {Large language models without grounding recover non-sensorimotor but not sensorimotor features of human concepts},
  journal = {Nature Human Behaviour},
  volume  = {9},
  number  = {9},
  pages   = {1871--1886},
  year    = {2025},
  doi     = {10.1038/s41562-025-02203-8}
}

@article{jiang2025codegen,
  author  = {Jiang, Juyong and Wang, Fan and Shen, Jiasi and Kim, Sungju and Kim, Sunghun},
  title   = {A Survey on Large Language Models for Code Generation},
  journal = {ACM Transactions on Software Engineering and Methodology},
  year    = {2025},
  doi     = {10.1145/3747588}
}

@article{moinuddin2021feedback,
  author  = {Moinuddin, Arsalan and Goel, Ashish and Sethi, Yashendra},
  title   = {The Role of Augmented Feedback on Motor Learning: {A} Systematic Review},
  journal = {Cureus},
  volume  = {13},
  number  = {11},
  pages   = {e19695},
  year    = {2021},
  doi     = {10.7759/cureus.19695}
}

@inproceedings{li2024iris,
  author        = {Li, Ziyang and Dutta, Saikat and Naik, Mayur},
  title         = {{IRIS}: {LLM}-Assisted Static Analysis for Detecting Security Vulnerabilities},
  booktitle     = {Proceedings of the International Conference on Learning Representations},
  year          = {2025},
  eprint        = {2405.17238},
  archivePrefix = {arXiv}
}

@article{moreschini2025techdebt,
  author  = {Moreschini, Sergio and others},
  title   = {The Evolution of Technical Debt from {DevOps} to Generative {AI}: {A} Multivocal Literature Review},
  journal = {Journal of Systems and Software},
  year    = {2025},
  doi     = {10.1016/j.jss.2025.112599}
}

@article{mesko2023llmregulation,
  author  = {Mesk{\'o}, Bertalan and Topol, Eric J.},
  title   = {The imperative for regulatory oversight of large language models (or generative {AI}) in healthcare},
  journal = {npj Digital Medicine},
  volume  = {6},
  pages   = {120},
  year    = {2023},
  doi     = {10.1038/s41746-023-00873-0}
}

@article{ebinger2025fdareg,
  author  = {Ebinger, Joseph and others},
  title   = {{United States Food and Drug Administration} Regulation of Clinical Software in the Era of Artificial Intelligence and Machine Learning},
  journal = {Mayo Clinic Proceedings: Digital Health},
  volume  = {3},
  number  = {3},
  pages   = {100231},
  year    = {2025},
  doi     = {10.1016/j.mcpdig.2025.100231}
}

@article{zaatar2024transformative,
  title={The transformative power of music: Insights into neuroplasticity, health, and disease},
  author={Zaatar, Muriel T and Alhakim, Kenda and Enayeh, Mohammad and Tamer, Ribal},
  journal={Brain, behavior, \& immunity-health},
  volume={35},
  pages={100716},
  year={2024},
  publisher={Elsevier}
}

@article{rebecchini2021music,
  title={Music, mental health, and immunity},
  author={Rebecchini, Lavinia},
  journal={Brain, behavior, \& immunity-health},
  volume={18},
  pages={100374},
  year={2021},
  publisher={Elsevier}
}

@article{schnall2016user,
  title={A user-centered model for designing consumer mobile health (mHealth) applications (apps)},
  author={Schnall, Rebecca and Rojas, Marlene and Bakken, Suzanne and Brown, William and Carballo-Dieguez, Alex and Carry, Monique and Gelaude, Deborah and Mosley, Jocelyn Patterson and Travers, Jasmine},
  journal={Journal of biomedical informatics},
  volume={60},
  pages={243--251},
  year={2016},
  publisher={Elsevier}
}

@inproceedings{kantan2024making,
  title={Making movement sonification usable in clinical gait rehabilitation: A user-centered study},
  author={Kantan, Prithvi Ravi and Dahl, Sofia and J{\o}rgensen, Helle Rovsing and Spaich, Erika G},
  booktitle={Proceedings of the 19th International Audio Mostly Conference: Explorations in Sonic Cultures},
  pages={43--60},
  year={2024}
}

@article{hamilton2019experiences,
  title={Experiences of therapists using feedback-based technology to improve physical function in rehabilitation settings: a qualitative systematic review},
  author={Hamilton, Caitlin and Lovarini, Meryl and McCluskey, Annie and Folly de Campos, Tarcisio and Hassett, Leanne},
  journal={Disability and Rehabilitation},
  volume={41},
  number={15},
  pages={1739--1750},
  year={2019},
  publisher={Taylor \& Francis}
}

@article{hamilton2022usability,
  title={Usability of affordable feedback-based technologies to improve mobility and physical activity in rehabilitation: a mixed methods study},
  author={Hamilton, Caitlin and Lovarini, Meryl and van den Berg, Maayken and McCluskey, Annie and Hassett, Leanne},
  journal={Disability and Rehabilitation},
  volume={44},
  number={15},
  pages={4029--4038},
  year={2022},
  publisher={Taylor \& Francis}
}

@book{izhaki2017mixing,
  title={Mixing audio: concepts, practices, and tools},
  author={Izhaki, Roey},
  year={2017},
  publisher={Routledge}
}

@book{huber2013modern,
  title={Modern recording techniques},
  author={Huber, David Miles and Runstein, Robert},
  year={2013},
  publisher={Routledge}
}

@inproceedings{kantan2026beyond,
  title={Beyond Deterministic Mappings: Audiovisual Correspondence in Movement-Controlled Generative Music},
  author={Kantan, Prithvi},
  booktitle={Proceedings of the 10th International Conference on Movement and Computing},
  pages={1--10},
  year={2026}
}

@article{alelyani2025aiframework,
  author  = {Alelyani, Turki},
  title   = {A validated framework for responsible {AI} in healthcare autonomous systems},
  journal = {Scientific Reports},
  volume  = {15},
  pages   = {44432},
  year    = {2025},
  doi     = {10.1038/s41598-025-25266-z}
}

@article{mienye2024xai,
  author  = {Mienye, Ibomoiye Domor and Obaido, George and Jere, Nobert and Mienye, Ebikella and Aruleba, Kehinde and Emmanuel, Ikiomoiye Douglas and Ogbuokiri, Blessing},
  title   = {A survey of explainable artificial intelligence in healthcare: {Concepts}, applications, and challenges},
  journal = {Informatics in Medicine Unlocked},
  volume  = {51},
  pages   = {101587},
  year    = {2024},
  doi     = {10.1016/j.imu.2024.101587}
}

@article{maes2016_3MO,
  author={Maes, Pieter-Jan and Buhmann, Jeska and Leman, Marc},
  year={2016},
  title={3MO: A Model for Music-Based Biofeedback},
  volume={1},
  journal={Frontiers in Neuroscience}
}

\end{document}